\documentclass[notitlepage,aps,prd,onecolumn,groupedaddress,showpacs,eqsecnum]{revtex4-1}
\usepackage[latin9]{inputenc}
\usepackage{amsmath}
\usepackage{amssymb}
\usepackage{graphicx}
\usepackage{esint}
\usepackage[usenames,dvipsnames]{xcolor}
\usepackage[colorlinks=true,linkcolor=blue,urlcolor=black,citecolor=blue]{hyperref}


\begin{document}

\title{Continuity of Scalar Fields With Logarithmic Correlations}

\author{S. G. Rajeev}

\email{s.g.rajeev@rochester.edu}

\altaffiliation[Also at the ]{Department of Mathematics}

\author{Evan Ranken}

\email{evan@pas.rochester.edu}

\affiliation{Department of Physics and Astronomy\\
 University of Rochester\\
 Rochester, New York 14627, USA}

\date{\today}
\begin{abstract}
We apply select ideas from the modern theory of stochastic processes
in order to study the continuity/roughness of scalar quantum fields.
A scalar field with logarithmic correlations (such as a massless field
in 1+1 spacetime dimensions) has the mildest of singularities, making
it a logical starting point. Instead of the usual inner product of
the field with a smooth function, we introduce a moving average on
an interval which allows us to obtain explicit results and has a simple
physical interpretation. Using the mathematical work of Dudley, we
prove that the averaged random process is in fact continuous, and
give a precise modulus of continuity bounding the short-distance variation.
% Referee edit 1
\end{abstract}

\pacs{ 02.50.-r, 11.10.Cd , 11.10.Kk, 84.37.+q }

\maketitle

\section{Introduction}

In traditional geometry, the distance between two points is the length
of the shortest curve that joins them. This fits well with classical
physics, as this shortest path is the one followed by a free particle.
But in quantum physics, the shortest one is only the most likely of
many paths that the particle can take. Moreover, no particle can follow
a path connecting two spacelike separated points. Taking these facts
into account, we should hesitate to associate distance with the length
of one particular curve. Instead, we can average over all paths connecting
two points, yielding the Green's function of a quantum field (also
called the two point function, correlation function or propagator.)
A metric does emerge out of the correlation, but turns out to be non-Euclidean
\cite{KarRajeevMetric}.

The idea of defining a metric from the correlation of a random process
is a staple of modern stochastic analysis \cite{Dudley,LedouxTalagrand,Adler,Bogachev}.
This can be illustrated with Brownian motion. The Brownian paths are
continuous, but not differentiable with respect to the usual time
parameter. A particle executing Brownian motion is knocked around
by other particles in the medium. As the time between collisions tends
to zero, the velocity at any instant is no longer a physical quantity.
Furthermore, even the speed cannot be bounded; as $x\to y$ the probability
of $\frac{|B(x)-B(y)|}{|x-y|}$ being bounded is zero (To make comparison
with a quantum field easier, we call the time parameter of the Brownian
process $x$ rather than $t$. Since the diffusion constant has dimension
$(\mathrm{length)}^{2}/\mathrm{time}$, dimensional analysis suggests
that 
\begin{eqnarray}
\frac{|B(x)-B(y)|}{\sqrt{|x-y|}}
\end{eqnarray}
would be a better quantity to measure the speed of a Brownian particle.
But it turns out that even this is unbounded with probability 1 (more
commonly stated as ``almost surely'' or a.s.) as $x\to y$. The
proper way to quantify the time that has elapsed between two measurements
is not $|x-y|$ or even $\sqrt{|x-y|}$. We seek a metric $\omega$
with respect to which the sample paths are locally Lipschitz continuous,
meaning $\frac{|B(x)-B(y)|}{\omega(x,y)}$ is almost surely bounded
as $x\rightarrow y$ . The correct such ``modulus of continuity,''
attributed to L$\mathrm{\acute{e}}$vy, is 
\begin{equation}
\omega(x,y)\propto\sqrt{|x-y|\log\frac{1}{|x-y|}}
\end{equation}
for small $|x-y|$. This quantifies the roughness of Brownian paths
(One can bound the variations precisely with a proportionality constant
$\sqrt{2}$, but we will generally ignore multiplicative constants
in discussing continuity/roughness here).

We look at the spatial metric in the simplest relativistic theory,
a massless scalar quantum field in $1+1$ dimensions. Such logarithmically
correlated fields have generated interest in purely mathematical contexts,
and have potential applications in areas ranging from finance to cosmology
(see \cite{LogCorrFields}). It is enough to understand the continuity
of sample fields in the ground state; those in any state of finite
energy will exhibit identical behavior over small distances (see Appendix
B for a discussion of the ground state wavefunction).

A complication is that scalar quantum fields are random distributions
rather than functions: $\phi(x)$ at some point in space is not a
meaningful quantity. But we will show that a mild smoothing procedure
(averaging over an interval) is enough to get around this difficulty,
yielding a continuous but not differentiable function. This average
can be viewed as a model for the potential measured by a device: such
a measurement will always take place over some finite width. A peculiar
property of the logarithmically correlated field is that the probability
law of the average is independent of the size of the interval (size
of the measuring device). That is, the field does not appear any rougher
if we average over smaller intervals.

We then obtain a result analogous to that of Levy: a metric in space
with respect to which the scalar field is a.s. Lipschitz (we will
use this term exclusively in the sense of local continuity). Our result
is a particular case of the much deeper mathematical theory of regularity
of random processes \cite{Dudley,LedouxTalagrand,Adler,Bogachev}
. The idea of using a moving average (instead of an inner product
with smooth test functions) seems to be new, and yields simple explicit
results.

We then apply this moving average technique to other random fields
of physical interest, noting that a new procedure is sometimes needed
if the field considered has more severe divergences. For supplemental
context, Appendix A discusses the intimate relationship of this work
to the resistance metric on a lattice, connecting to an earlier paper
\cite{KarRajeevMetric}, while Appendix B describes connections to
a functional analytic approach to regularity of random processes.

Although we work with Gaussian fields in this paper, the short distance
behavior is the same for asymptotically free interacting fields (up
to sub-leading logarithmic corrections). The regularity of renormalizable
but not asymptotically-free theories (such as QED or the Higgs model)
can be quite different. The strength of interactions grow as distances
shrink, possibly leading to a singularity (Landau pole). In the case
of QED, we know that this is not physically significant, due to unification
with weak interactions into a non-Abelian gauge theory.

But the question of what happens to the self-interaction of a scalar
quantum field (Higgs boson) at short distances is still open. In the
absence of evidence at the LHC for supersymmetry or compositeness
of the scalar, we have to consider the possibility that the Higgs
model is truly the fundamental theory. The short distance behavior
is dominated by interactions, necessitating new mathematical methods
beyond perturbative renormalization. The extensive mathematical literature
\cite{Dudley,LedouxTalagrand,Adler,Bogachev} on continuity of non-Gaussian
processes ought to contain useful tools for physics. In order to apply
this work to a full interacting theory, we must first know what happens
in the simpler case of a free theory. This is part of the physical
motivation for this paper.

\section{Continuity }

\subsection{Continuity of Random Processes}

A random process $r(x)$ assigns a random variable to each value of
$x$ in some space $X$. The quantity

\begin{equation}
d(x,y)=\sqrt{\langle\left[r(x)-r(y)\right]^{2}\rangle}\label{dxy}
\end{equation}

satisfies the triangle inequality and so defines a metric on $X$
(provided we identify any originally distinct points $x,y$ for which
$d(x,y)=0$).

This metric need not be Euclidean or even Riemannian. A standard example
is Brownian motion, where $d(x,y)=\sqrt{|x-y|}$, which is neither.
We work out another simple case in Appendix A: when $X$ is a finite
graph, and $d$ is the square root of the resistance metric \cite{ResistanceMetric,resistance2,resistance3}.

One commonly successful approach to the study of continuity is to
leave behind the intuitive structure associated with the space $X$,
and begin instead by looking at structures related to the process
of interest (such as the metric $d$ above). At first, one might expect
that the sample paths for a random process $r(x)$ will be necessarily
continuous with respect to $d$. Although true for Brownian motion,
almost sure continuity with respect to $d$ does not hold in general.
For Gaussian processes, a sufficient condition for continuity is the
convergence of the Dudley integral \cite{Dudley,LedouxTalagrand,Adler,Bogachev}

\begin{equation}
J(\delta)=\int_{0}^{\delta}\sqrt{\log N(D,\epsilon)}\,d\epsilon,\quad\delta<D
\end{equation}

% Referee edit 2where $N(D,\epsilon)$ be the minimum number of balls
of radius $\epsilon$ it takes to cover a ball of radius $D$ in $(X,d)$
(We suppress the $D$ dependence of $J$ for simplicity of notation).
The possible divergence comes from the lower limit of the integral
$\epsilon\rightarrow0$.

Can we go beyond continuity? To speak of differentiable functions,
a metric is not enough: we would need a differentiable structure on
$X$ which we do not have intrinsically. The closest analogue to differentiable
functions on a metric space $(X,d)$ are Lipschitz functions, for
which $\frac{|f(x)-f(y)|}{d(x,y)}$ is bounded. For comparison, differentiable
functions on the real line are Lipschitz, but not all Lipschitz functions
are differentiable. Of course, all Lipschitz functions are continuous.

Even in cases where $J(\delta)$ converges, indicating that the sample
paths are continuous, they may still not be Lipschitz with respect
to the metric $d$ above. Again, the Dudley integral comes to the
rescue: using it we can define a more refined metric

\begin{equation}
\omega(x,y)=J(d(x,y)).
\end{equation}
% Referee edit 3Since $\sqrt{\log N(D,\epsilon)}$ is a decreasing
function of $\epsilon$, $J(\delta)$ is a convex function. Thus $J(d(x,y))$
satisfies the triangle inequality as well.

The sample paths of a Gaussian process for which $J(\delta)$ converges
are \cite{Dudley} a.s. Lipschitz in this refined metric $\omega$.
Thus $\omega$, rather than $d$, is the metric (``modulus of continuity'')
we must associate to a Gaussian random process.

What would one do if the Dudley integral does not converge? There
is a more general theory \cite{LedouxTalagrand} which gives necessary
\emph{and sufficient} conditions for continuity: a ``majorizing measure''
must exist on $X$ . We will not use this theory in this paper, but
hope to return to it, as it can deal with more general cases than
Gaussian processes (e.g., interacting quantum fields).

\subsection{Quantum Fields}

In this paper, we consider a quantum scalar field $\phi$ in the continuum
limit. In the trivial case where $\phi$ is massless with 1 spatial
dimension and no time dimension, the correlations of Brownian motion
are reproduced and $\phi$ remains a continuous function. However,
in any fully relativistic field theory, $\phi$ lives on a space of
distributions, not functions. To get a sensible random variable, we
must then take the inner product with respect to some test function
$h$ with zero average.

\begin{equation}
\phi[h]=\int\phi(x)h(x)dx,\quad\int h(x)dx=0.
\end{equation}

We study the case where $\phi$ is a distribution with the weakest
possible singularities; one might say we want a field that is ``close''
to being a function. The obvious candidate is the case of logarithmic
correlations (For a recent review, see \cite{LogCorrFields})

\begin{equation}
\langle\phi[h]\phi[h']\rangle=-\int\log|x-y|\ h(x)h'(y)dxdy.
\end{equation}

The corresponding Gaussian measure can be thought of as the (square
of the) ground state wavefunction of a massless scalar field in $1+1$
dimensions. (More precisely, the continuum limit of the resistance
metric of a row on an infinite square lattice, discussed in Appendix
A).

The condition $\int h(x)dx=0$ ensures that the covariance is unchanged
if $\log|x-y|$ is replaced by $\log\lambda|x-y|$, meaning $\phi$
is scale invariant. Since $\phi$ has the physical meaning of a potential,
observables such as $\phi[h]$ must be unchanged under a shift $\phi(x)\mapsto\phi(x)+a$,
which equivalently suggests the requirement $\int h(x)dx=0$.

\subsection{Moving Average of a Quantum Field}

Quantum fields which are only mildly singular can act on test functions
which are not smooth or even continuous. It is not necessary to consider
the whole space of test functions as in \cite{Adler}; in this paper
our test functions will be piecewise constant with compact support
and zero mean.

We define a \emph{moving average} of $\phi$: 
\begin{equation}
\bar{\phi}_{s}(u)\;=\;\int_{-\frac{1}{2}}^{\frac{1}{2}}\left\{ \phi\left(s\left[u-w\right]\right)-\phi\left(s\left[0-w\right]\right)\right\} dw,\quad s>0.
\end{equation}
This is the inner product of $\phi$ with a discontinuous test function
$h$ that has support on two intervals of width $s$ based at $su$
and at $0$; the sign is chosen so that $\int h(x)dx=0$. The probability
law of $\bar{\phi}_{s}$ is not translation invariant: the second
term ensures the boundary condition 
\begin{equation}
\bar{\phi}_{s}(0)=0.
\end{equation}

%Referee edit 4We will mostly work with the quantity 
\begin{equation}
\bar{\phi}_{s}(u)-\bar{\phi}_{s}(v)\;=\;\int_{-\frac{1}{2}}^{\frac{1}{2}}\left\{ \phi\left(s\left[u-w\right]\right)-\phi\left(s\left[v-w\right]\right)\right\} dw
\end{equation}
which has a translation invariant law. It is convenient to rescale
the coordinate of the midpoint by the width (as we have already done),
so that the variable $u$ is dimensionless. Then the quantity

\begin{equation}
\rho(u,v)\equiv\sqrt{\langle\left[\bar{\phi}_{s}(u)-\bar{\phi}_{s}(v)\right]^{2}\rangle},
\end{equation}

which is just a special case of (\ref{dxy}), is finite and defines
a metric. Moreover, it is independent of $s$ in the logarithmically
correlated case. This means the process $\bar{\phi}_{s}(u)-\bar{\phi}_{s}(v)$
has a probability law that is independent of $s$: a consequence of
scale invariance, which is specific to logarithmic correlations. As
an interesting aside, we note that $s\bar{\phi}_{s}(u)$ produces
a solution to the wave equation in $u$ and $s$.

The moving average does not depart from the essence of the standard
idea of averaging over a test function. It is simply that a piecewise
constant test function is especially convenient for a mildly singular
quantum field as opposed to a smoother function. For more singular
fields (e.g. scalar field in four dimensions) we would have to revert
to more regular test functions.

\section{Logarithmically correlated scalar field in 1 dimension}

In the logarithmically correlated case, we obtain explicit formula

\begin{equation}
\rho(u,v)=\rho(|u-v|)
\end{equation}

\begin{equation}
\rho(r)=\sqrt{L(r+1)+L(r-1)-2L(r}),
\end{equation}

Where

\begin{equation}
L(r)=\frac{1}{2}r^{2}\log r^{2}.\label{Ldef}
\end{equation}

\begin{figure}
\begin{minipage}[c][1\totalheight][t]{0.45\textwidth}%
\includegraphics[width=1\textwidth]{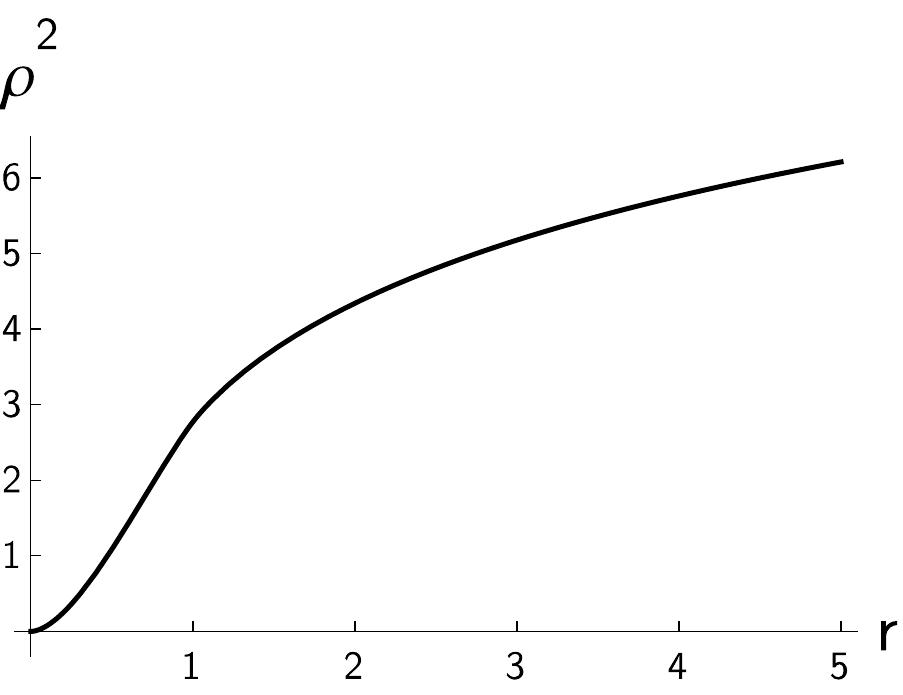}%
\end{minipage}\hfill{}%
\begin{minipage}[c][1\totalheight][t]{0.48\textwidth}%
\includegraphics[width=1\textwidth]{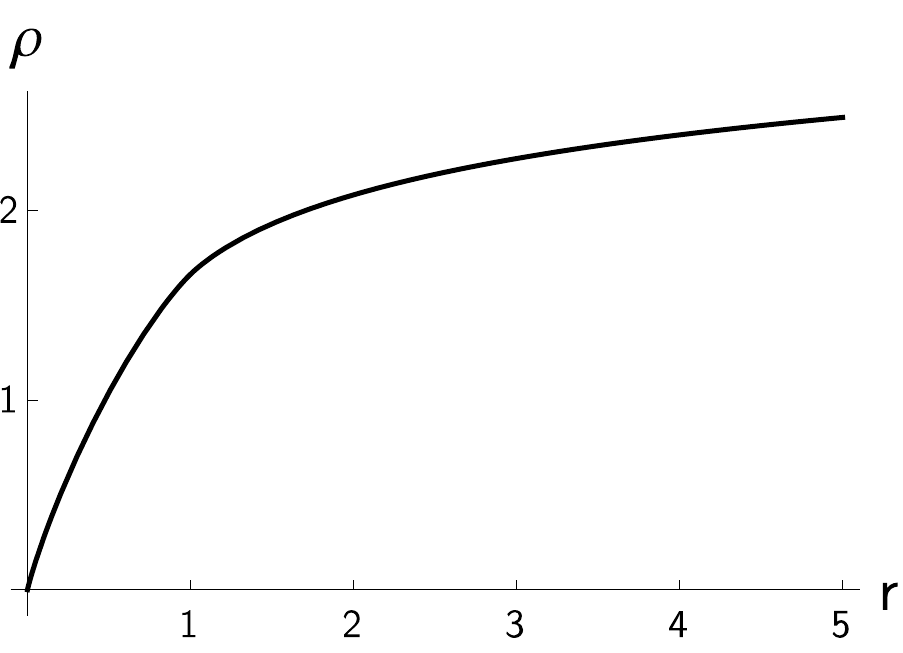}%
\end{minipage}

\protect\protect\protect\protect\protect\caption{The behavior of $\rho^{2}(r)$ and $\rho(r)$, setting $s=1$. \label{fig:rho}}
\end{figure}

Being a convex function of $r=|u-v|$, this $\rho(u,v)$ will satisfy
the triangle inequality (not true of $\rho^{2}$, as seen in Fig.
\ref{fig:rho}). Thus, $\rho$ defines a translationally invariant
metric.

Simple calculations (see Sec. \ref{sec:Calculations-and-further}) show
that the Dudley integral $J$ converges, so that $\bar{\phi}_{s}$
is a.s. continuous in $\rho$. Moreover we can construct a refinement

\begin{equation}
J\left(\rho(u,v)\right)\equiv\omega(u,v)\approx|u-v|\log\frac{1}{|u-v|}
\end{equation}

%Referee edit 5with respect to which $\bar{\phi}_{s}$ is a.s. Lipschitz.
This is a ``modulus of continuity'' for the moving average of a
quantum field, analogous to that of L$\mathrm{\acute{e}}$vy for Brownian
motion. (Note that there is no square root, however.)

\begin{figure}
\begin{minipage}[c][1\totalheight][t]{0.45\textwidth}%
\includegraphics[width=1\textwidth]{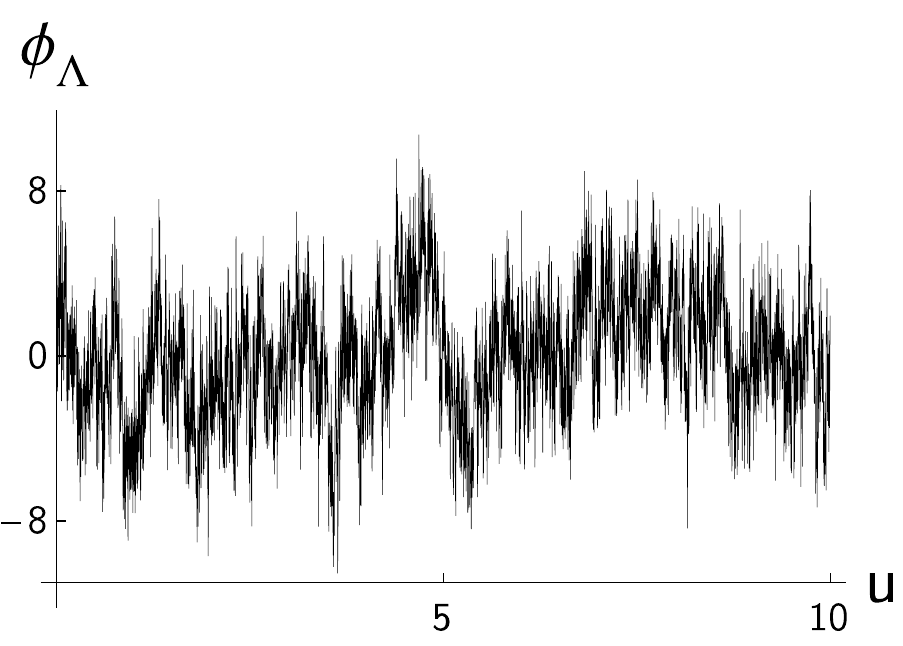}%
\end{minipage}\hfill{}%
\begin{minipage}[c][1\totalheight][t]{0.45\textwidth}%
\includegraphics[width=1\textwidth]{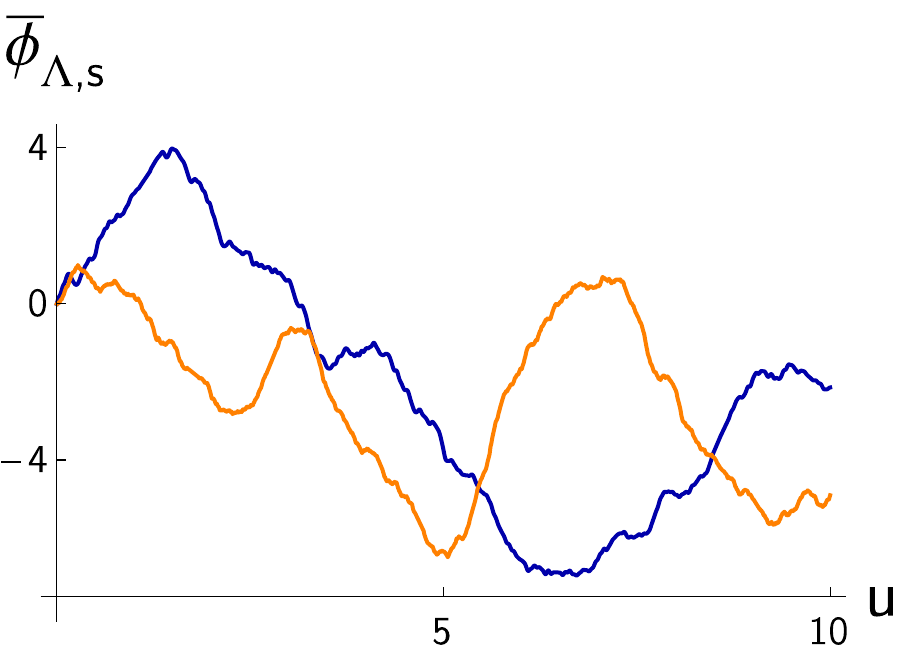}%
\end{minipage}

\protect\protect\protect\protect\protect\caption{We show an approximate $\phi$ and two averages $\bar{\phi}_{\Lambda,s}$
where $\Lambda=4000$, $L=5$ and we average over width $s=0.05$
(blue/dark) and $s=0.2$ (orange/light). We see that despite the factor
of 4 difference in averaging windows, the two appear interchangeable,
demonstrating the scale invariance even for this approximate representation.
The two appear to follow the same law, and the continuity is seen
to be greatly improved. \label{fig:noise}}
\end{figure}

We can obtain a crude picture of the moving average process by generating
noise which has the same power spectrum as a log-correlated field,
but with some high frequency cutoff. This is given by the Fourier
series %Referee edit 6
\begin{equation}
\phi_{\mathrm{_{\Lambda}}}(x)=\sum_{m=1}^{\Lambda}\frac{1}{\sqrt{m}}\left[X_{m}\cos\left(\frac{\pi mx}{L}\right)+Y_{m}\sin\left(\frac{\pi mx}{L}\right)\right]
\end{equation}

where $X_{m},Y_{m}$ are independent standard Gaussian variables.
For large $\Lambda$ (ultraviolet cutoff) and $L$ (the infrared cutoff,
$-L<x<L$ ) this creates an intuitive ``approximation'' to the divergent
field. Such a technique is often used to visualize white noise. While
one must be careful claiming to ``approximate'' a distribution with
a truncated series, $\phi_{\mathrm{_{\Lambda}}}$ gives us some sensible
object on which to numerically test the properties of our moving average.
This is carried out in Fig. \ref{fig:noise}. We see that $\bar{\phi}_{s}$
has the desired properties without requiring the full distribution.

While we focus on the log-correlated field for its mathematical simplicity,
it is worth noting that such objects are not necessarily confined
to the realm of mathematical fantasy. A free scalar field in one dimension
can in principle be a good approximation for a real physical system,
with one possible example being the electromagnetic field of certain
optical fibers. If the refractive index of the fiber is chosen appropriately,
only a finite number of transmission modes will be allowed. We can
think of the wave equation as analogous to the Schrödinger equation,
with the variable refractive index providing an effective potential.
This potential can be chosen to allow only a finite number of bound
states. Single-mode fibers have only one such state, leading to a
system with one effective spatial dimension.

Even in the absence of light in the fiber (ground state of the electromagnetic
field), there will be quantum fluctuations in the potential. In the
absence of severe nonlinearities, these fluctuations can be modeled
as two noninteracting scalar fields, one for each polarization mode.
If the wire is transparent over a sufficient frequency (maintaining
its single-mode property and minimal dispersion for propagating waves),
then the potential difference between two points will be a Gaussian
random variable whose variance is approximately logarithmic with distance.
The measurement of the potential would require a probe of finite size,
so the averaging process employed in this paper provides a convincing
model for the potential as seen by a measuring apparatus at a given
instant. The considerations of this paper can be viewed as a model
of the spatial regularity of the electromagnetic potential in such
an optical fiber. This model could also, in principle, describe the
ground state fluctuations of a quantum system confined to a very narrow
region in 2 spatial dimensions, sometimes called a quantum wire.

Perhaps an experimental test of the sample field behavior in Fig.
\ref{fig:noise} is indeed possible. However, the details of realizing
such a system and carrying out such measurements is highly nontrivial
and not suited to the themes of this paper; we include this discussion
mainly as a reminder that lower dimensional systems are often not
so unphysical as they seem.

\section{Explicit calculations and further examples\label{sec:Calculations-and-further}}

\subsection{Variance of $\bar{\phi}_{s}$ for the log-correlated field}

The calculations that justify the above assertions are straightforward,
but worth outlining as they help illuminate the properties discussed
above. Because of the divergences, we cannot use the standard approach
directly to the quantum field, but only to its moving average. Begin
with the observation that

\begin{equation}
F(a,b,c,d)\equiv-\int_{a}^{b}dx\int_{c}^{d}dy\log|x-y|\label{Fdef}
\end{equation}

\begin{equation}
=\frac{3}{2}(a-b)(c-d)+\frac{1}{2}\left[L(c-a)-L(d-a)-L(c-b)+L(d-b)\right],
\end{equation}

where $L$ is defined in (\ref{Ldef}). Note this quantity is not
quite scale invariant: there is an ``anomaly'' proportional to $\log\lambda$.

\begin{equation}
L(\lambda r)=\lambda^{2}L(r)+r^{2}\log\lambda
\end{equation}

\begin{equation}
F\left(\lambda a,\lambda b,\lambda c,\lambda d\right)=\lambda^{2}F(a,b,c,d)-(b-a)(d-c)\log\lambda
\end{equation}

Then

\begin{equation}
\langle\left[\bar{\phi}_{s}(u)-\bar{\phi}_{s}(v)\right]^{2}\rangle=\frac{F(a,b,a,b)}{(b-a)^{2}}+\frac{F(c,d,c,d)}{(d-c)^{2}}-2\frac{F(a,b,c,d)}{(b-a)(d-c)},
\end{equation}

where

\begin{equation}
\frac{F(a,b,a,b)}{(b-a)^{2}}=\frac{3}{2}-\frac{1}{2}\log(b-a)^{2}
\end{equation}

which only depends on the width of the interval $[a,b]$. We can then
consider two intervals of equal width $s$, centered at $su$ and
$sv$, yielding

\begin{equation}
\langle\left[\bar{\phi}_{s}(u)-\bar{\phi}_{s}(v)\right]^{2}\rangle=3-2\log(s)-\frac{2}{s^{2}}F\left(su-\frac{s}{2},su+\frac{s}{2},sv-\frac{s}{2},sv+\frac{s}{2}\right)
\end{equation}

From the scale transformation property above of $F$ we can see that
this quantity is independent of $s$: the ``scale anomaly'' of $F$
cancels against $2\log s$. So we can simplify by putting $s=1$ and
expressing $F$ in terms of $L$:

\begin{equation}
\begin{split}\langle\left[\bar{\phi}_{s}(u)-\bar{\phi}_{s}(v)\right]^{2}\rangle & \equiv\rho^{2}(u,v)\\
 & =L(u-v-1)+L(u-v+1)-2L(u-v)
\end{split}
\end{equation}

as was claimed.

\subsection{Continuity of Brownian Paths}

In using the Dudley integral, it is useful to begin with a well-known
example. The most familiar example of a Gaussian process is Wiener's
model of Brownian motion, for which $d(x,y)=\sqrt{|x-y|}$ . If an
interval $[0,1]$ is divided into $N$ equal parts, each part is contained
in a $d-$ball of radius $\epsilon=\sqrt{\frac{1}{2N}}.$ Thus $N(\epsilon)=1+\mathrm{Floor}\left(\frac{1}{2\epsilon^{2}}\right)$
and for small $\delta$, {[}where $\mathrm{Floor}(a)$ is the integer
part of the real number $a${]} %Referee edit 7

\begin{equation}
J(\delta)\approx\delta\sqrt{-2\log\delta}.
\end{equation}

Thus Brownian sample paths $B$ are almost surely continuous. More
quantitatively, we may construct

\begin{equation}
\omega(r)=J(d(r))=\sqrt{r\log(1/r)}.\label{omegaBrownian}
\end{equation}

to obtain the result of L$\mathrm{\acute{e}}$vy that, with probability
one,

\begin{equation}
\frac{|B(x)-B(y)|}{\sqrt{|x-y|\log\frac{1}{|x-y||}}}<C
\end{equation}

as $x\rightarrow y$ for some constant $C$.

\subsection{Continuity of $\bar{\phi}_{s}$ for logarithmically correlated fields}

We can now show that the sample paths $\bar{\phi}_{s}(u)$ are continuous
with probability one. Again, if $[0,1]$ is divided into $N$ intervals,
each will have radius $\epsilon=\rho\left(\frac{1}{N}\right)$. To
get small $\epsilon,$ we must choose a large $N$; using the asymptotic
behavior

\begin{equation}
\rho(r)\approx r\sqrt{-\log r}
\end{equation}

for small $r$,

\begin{equation}
\epsilon\approx\frac{1}{N}\sqrt{-\log\left[\frac{1}{N}\right]}\;\;\Longrightarrow\;\; N(\epsilon)\approx\frac{1}{\epsilon}\sqrt{-\log\epsilon}.
\end{equation}

The Dudley integral converges:

\begin{equation}
J(\delta)\approx\delta\sqrt{\log\frac{1}{\delta}},\quad\delta\to0
\end{equation}

\begin{equation}
J\left(\rho(r)\right)\equiv\omega(r)\approx r\log(1/r)\label{omegaLogcorr}
\end{equation}

which yields the claimed modulus of continuity.

\subsection{Additional Examples for Comparison}

\subsubsection{Moving Average of Brownian Paths}

It is informative to apply the moving average procedure to the Brownian
case, where the paths $B(x)$ which we average over are continuous
functions to begin with. Proceeding analogously, consider two intervals
with width $s$ with centers $su$ and $sv$ respectively. Then we
can define, analogous to (\ref{Fdef}) but with some added foresight,
\begin{equation}
F\left(su-\frac{s}{2},su+\frac{s}{2},sv-\frac{s}{2},sv+\frac{s}{2}\right)\equiv-\int_{su-s/2}^{su+s/2}dx\int_{sv-s/2}^{sv+s/2}dy|x-y|
\end{equation}
\begin{equation}
=\begin{cases}
\frac{1}{3}\left(r^{3}\left(-3s^{2}+3s-1\right)+3r^{2}s^{3}+s^{3}\right) & 0<r<s\\
s^{3}r & r>s,
\end{cases}
\end{equation}

where $r=|u-v|$. We then have
\begin{eqnarray*}
\langle\left[\bar{B}_{s}(u)-\bar{B}_{s}(v)\right]^{2}\rangle & \equiv & \rho^{2}(r)\\
 & = & \frac{2s}{3}-\frac{2}{s^{2}}F\left(su-\frac{s}{2},su+\frac{s}{2},sv-\frac{s}{2},sv+\frac{s}{2}\right).
\end{eqnarray*}

It is easily seen that $\rho^{2}(\lambda r)=\lambda\rho^{2}(r)$,
breaking scale invariance. Still for comparison purposes, we consider
averaging over intervals of width $s=1$, noting that the scaling
behavior will only change $\omega(r)$ by a constant factor.

As before, $\rho^{2}$ does not define a metric, but its square root
$\rho$ does. In the large-$r$ limit we have

\begin{equation}
\rho(r)\sim\sqrt{r},
\end{equation}

while for small $r$,

\begin{equation}
\rho(r)\sim r.
\end{equation}

This short distance behavior suggests by dimensional analysis that
$\bar{B}_{s}(x)$ might be Lipschitz in the usual metric $|u-v|$,
but the Dudley integral yields a weaker limit 
\begin{equation}
\omega(r)\approx r\log\left(\frac{1}{r}\right)
\end{equation}

\begin{equation}
\frac{|\bar{B}_{s}(u)-\bar{B}_{s}(v)|}{r\log(\frac{1}{r})}<C.
\end{equation}

Thus $\bar{B}_{s}$ is just shy of being Lipschitz in the usual metric,
but is a.s. Lipschitz with respect to the metric $\omega(u,v)\approx|u-v|\log\frac{1}{|u-v|}$.
Interestingly, this is the same $\omega$ we obtained in (\ref{omegaLogcorr})
for the log-correlated case, even though the short distance behavior
of $\rho$ is not quite the same (the difference in the Dudley integral
vanishes for small $\delta$). However $\omega$ for the Brownian
sample paths prior to averaging (\ref{omegaBrownian}) contains a square
root not present here.

\subsubsection{Power Law Correlations in 1D}

We can use the same method as with Brownian motion to consider the
moving average of a more general power-law correlated field such that
\begin{equation}
\langle\phi[h]\phi[h']\rangle=\mathrm{sign}(\alpha)\int|x-y|^{\alpha}h(x)h'(y)dxdy,\quad\quad\int h(x)dx=0.
\end{equation}

When $\alpha>0$ this is related to fractional Brownian motion \cite{fBM}.
When $\alpha=-1$ it is the restriction to one dimension of a massless
scalar quantum field in $2+1$ dimensions. The moving average is no
longer independent of the width of the intervals. Still, for purposes
of comparison, we consider the average on intervals of fixed width
$s=1$.

It is not difficult to evaluate the integrals to find that, in the
small $r$ limit,

\begin{equation}
\rho^{2}(r)\sim\begin{cases}
r^{2} & \alpha>0\\
r^{\alpha+2} & -2<\alpha<0,\;\alpha\neq-1\\
r\log r & \alpha=-1.
\end{cases}
\end{equation}

The moving average is Lipschitz with respect to the modulus

\begin{equation}
\omega(r)=\begin{cases}
r\log(1/r) & \alpha>0\\
r^{\frac{\alpha}{2}+1}\log(1/r) & -2<\alpha<0.
\end{cases}
\end{equation}

%Referee edit 8When $\alpha\leq-2$ the divergences are such that
the moving average is an insufficient tool to smooth the quantum field.
Note that $\omega(r)$ is the same in the logarithmic case as the
case where $\alpha>0$. The logarithmic case can be thought of as
the critical case where the smoothness implied by Dudley's criterion
starts to lessen.

\subsubsection{Log Correlated Scalar Field in 3D}

It is useful to work out a case in higher dimensions as well. The
massless scalar field in $n+1$ space-time dimensions has correlation

\begin{equation}
\big\langle\phi(x)\phi(y)\big\rangle\propto\frac{1}{|x-y|^{n-1}}\,.
\end{equation}

Thus for $n>1$ will we get power law, instead of logarithmic correlations.
Yet a logarithmically correlated, nonrelativistic, scalar field in
3 space dimensions is still of interest in cosmology \cite{LogCorrFields,Cosmology}.
As with the log-correlated scalar field in 1D we must average it over
a test function

\begin{equation}
\langle\phi[h]\phi[h']\rangle=-\int\log|x-y|h(x)h'(y)dx^{3}dy^{3},\quad\quad\int h(x)dx^{3}=0.
\end{equation}

Recall that

\begin{equation}
-\log|x|+\mathrm{const}=c\int e^{ik\cdot x}\frac{1}{|k|^{3}}\frac{d^{3}k}{(2\pi)^{3}}
\end{equation}

where the constant $c=2\pi^{2}$. The integral is not absolutely convergent,
so we define it through zeta regularization.

We perform our moving average over the interior of a sphere with radius
$t$, centered at $tu$

\begin{equation}
\bar{\phi}_{t}(u)\equiv\int_{|w|\leq1}\left\{ \phi\left(t\left[u-w\right]\right)-\phi\left(t\left[0-w\right]\right)\right\} dw
\end{equation}

\begin{equation}
\langle\bar{\phi}_{t}(u)\bar{\phi}_{t}(v)\rangle=\int_{|w|\leq1}\big\langle\phi\left(t\left[u-w_{1}\right]\right)\phi\left(t\left[v-w_{1}\right]\right)\big\rangle dw_{1}dw_{2}
\end{equation}

\begin{equation}
=c^{2}\int\frac{1}{|k|^{3}}e^{ik\cdot(u-v)t}\frac{d^{3}k}{(2\pi)^{3}}\int_{|w|\leq1}e^{-ik\cdot(w_{1}-w_{2})t}dw_{1}dw_{2}.
\end{equation}

Taking $t=1$, this can be reduced to the form

\begin{equation}
\langle\left[\bar{\phi}_{t}(u)-\bar{\phi}_{t}(v)\right]^{2}\rangle=2^{6}\pi^{4}G(r)
\end{equation}

where

\begin{equation}
G(r)=\int_{0}^{\infty}dk\frac{1}{k^{7}}\left[\sin k-k\cos k\right]^{2}\left[1-\frac{\sin kr}{kr}\right]
\end{equation}

and $r=|u-v|$. We are not able to evaluate the integral analytically,
but its convergence is clear, justifying the independence on $t$.
In the large $r$ limit, the integral is dominated by small $k$ contribution.
We then have

\begin{equation}
G(r)\approx\int_{0}^{\infty}dk\frac{1}{3k}\left[1-\frac{\sin kr}{kr}\right]
\end{equation}

\begin{equation}
\sim\log(r)+O(1).
\end{equation}

In the case of small $r$, the dominant contribution comes from the
first peak of $\frac{1}{k^{7}}\left[\sin k-k\cos k\right]^{2}$, which
must occur for $k<2\pi$ (i.e., $k\approx5.678$). This allows us
to treat $kr$ as small, yielding the behavior 
\begin{equation}
G(r)\approx r^{2}\int_{0}^{\infty}dk\frac{1}{3k^{5}}\left[\sin k-k\cos k\right]^{2} \;
\sim r^{2}.
\end{equation}

The approximation can be verified numerically. This small $r$ behavior
dictates the continuity modulus discussed above. Namely, we have that
for the 3D log-correlated scalar field, 
\begin{eqnarray}
\rho(r)&\sim &  r, \\
\omega(r)&\sim &  r\log(1/r).
\end{eqnarray}

A similar metric can be obtained for a log correlated field in other
dimensions. Note that, once we have $\rho(r)\sim r$ for small $r$,
the logarithm in the Dudley integral ensures that $\omega$ will not
depend on the dimensionality (up to proportionality). This is not
true if $\rho(r)$ has some other short distance behavior.

\section{Outlook}

Gaussian processes correspond to free fields. The most elegant way
to introduce interactions into a scalar field theory is to let it
take values in a curved Riemannian manifold. This is the nonlinear
sigma model in physics language, or the wave map in the mathematical
literature. In 1+1 dimensions, such a theory, with a target space
of a sphere or a compact Lie group, is well studied in the physics
literature. The short distance behavior is approximated by free fields
with corrections computable in perturbation theory (asymptotic freedom).
The only case for which mathematically rigorous results are known
is that of the Wess-Zumino-Witten model, which has non-Gaussian behavior
at short distances; i.e., a ``nontrivial fixed point'' for the renormalization
group. The related measure for the ground state of the quantum field
has been constructed by Pickrell. (For a review, see \cite{Pickrell}).
It is natural to ask for regularity results analogous to ours in this
case.

Looking further out, it would be interesting to quantify the regularity
of quantum fields of the nonlinear sigma model in two dimensional
space time; and even further out, $\lambda\phi^{4}$ theory in four
dimensions. It is possible that the ``naturalness problem'' of the
standard model of particle physics has a resolution in terms of such
a deeper understanding of the regularity of scalar quantum field theory.
The ``modern'' theory \cite{LedouxTalagrand} of regularity of non-Gaussian
processes ought to help with this daunting task. Even harder is the
case of Yang-Mills fields. An analogue of our moving average is the
Wilson loop. The measure of integration over the space of gauge fields
is only known rigorously for the two dimensional case \cite{Sengupta}.
Regularity of Yang-Mills fields satisfying classical evolution equations
(let alone random processes) is already a formidable problem under
active investigation (see for example \cite{YM3}).

\section{Acknowledgements}

We thank L. Gross, A. Iosevich, A. Jordan, C. Mueller and D. Pickrell
for discussions. The work of E.R. is supported by the NSF Graduate
Research Fellowship Program and a University of Rochester Sproull
Fellowship.

\appendix
%dummy comment inserted by tex2lyx to ensure that this paragraph is not empty

\section{The Resistance Metric as the Variance of Potential Fluctuations \label{sec:Appendix:-resistance-metric}}

Without being aware of the ``modern'' theory \cite{Adler,Bogachev,Dudley,LedouxTalagrand}
of random processes, we argued in an earlier paper \cite{KarRajeevMetric}
that the two point function (for spacelike separations) of a quantum
scalar field

\begin{equation}
\sqrt{\langle\left[\phi(x)-\phi(y)\right]^{2}\rangle}
\end{equation}

be used as the metric on spacetime. Since quantities such as $<\phi^{2}(x)>$
are divergent in a quantum field theory, the metric was defined with
a regularization. With the lattice regularization of a free massless
scalar field, our proposal for the metric fitted well with the idea
of a resistance metric \cite{ResistanceMetric,resistance2,resistance3}
popular in network theory.

In this appendix we show that the resistance metric (more precisely
its square root) is simply a finite dimensional special case of the
metric $d$ appearing in the theory of Gaussian processes. This connection
can be thought of as a particular case of the fluctuation-dissipation
theorem of statistical mechanics: the potential difference across
a resistor has thermal fluctuations with variance proportional to
dissipation.

Imagine each edge of a network as a unit resistor connecting two vertices.
Then, if a unit potential difference is applied across two vertices
$(k,l$), the reciprocal of the power dissipated defines the effective
resistance $R_{kl}$ between them. Kirchhoff's laws imply a variational
principle for this quantity \cite{ResistanceMetric}

\begin{equation}
R_{kl}=\frac{1}{\inf_{\phi}\left\{ \sum_{ij}A_{ij}(\phi_{i}-\phi_{j})^{2}\mid\phi_{k}-\phi_{l}=1\right\} }
\end{equation}

where $A$ is the adjacency matrix of the network. It is well known
that this $R_{kl}$ satisfies the triangle inequality, and is used
as a metric in network theory.

It is convenient to introduce another symmetric matrix $K$ by

\begin{equation}
\sum_{ij}A_{ij}(\phi_{i}-\phi_{j})^{2}=\sum_{ij}K_{ij}\phi_{i}\phi_{j}.
\end{equation}

\subsection{Some Linear Algebra}

%Referee edit 9Note that $K$ is degenerate; it vanishes on the vector
whose components are all equal to one:

\begin{equation}
\sum_{j}K_{ij}c_{j}=0,\quad c\equiv\left(1,1,,\cdots1\right).
\end{equation}

In particular, the equation

\begin{equation}
\sum_{j}K_{ij}\phi_{j}=J_{i}
\end{equation}
has a solution only if

\begin{equation}
\sum_{i}J_{i}=0.
\end{equation}

But the solution is not unique because if $\phi_{i}$is a solution,
so is $\phi_{i}+ac_{i}$ .We can construct an inverse for $K$ by
restricting the potentials to the subspace satisfying

\begin{equation}
\sum_{i}\phi_{i}=0.
\end{equation}

This fixes the overall constant (``ground potential'') in $\phi_{i}$.
Now, $K$ is an invertible map of this $n-1$ dimensional subspace
to itself; there is a matrix $G$ satisfying

\begin{equation}
\phi_{i}=\sum_{j}G_{ij}J_{j}.
\end{equation}

Equivalently, we can define $G$ by the equations

\begin{equation}
\sum_{i}G_{ij}=0
\end{equation}

\begin{equation}
\sum_{j}K_{ij}G_{jk}=\delta_{ik}-\frac{1}{n}c_{i}c_{k}.
\end{equation}

\subsection{Variational Principle}

We can solve this variational problem for effective resistance using
a Lagrange multiplier:

\begin{equation}
S=\sum_{ij}A_{ij}(\phi_{i}-\phi_{j})^{2}+\lambda(\phi_{k}-\phi_{l})
\end{equation}

\begin{equation}
\frac{\partial S}{\partial\phi_{i}}=0\implies2\sum_{j}K_{ij}\phi_{j}+\lambda\left[\delta_{ik}-\delta_{il}\right]=0.
\end{equation}

The solution is

\begin{equation}
\phi_{i}=-\frac{\lambda}{2}\left[G_{ik}-G_{il}\right].
\end{equation}

The constraint $\phi_{k}-\phi_{l}=1$ determines $\lambda$:

\begin{equation}
\lambda=-\frac{2}{[G_{kk}+G_{ll}-G_{kl}]}\,.
\end{equation}

Then

\begin{equation}
\sum K_{ij}\phi_{i}\phi_{j}=-\frac{\lambda}{2}\sum_{i}\phi_{i}(\delta_{ik}-\delta_{il})=-\frac{\lambda}{2}(\phi_{k}-\phi_{l})
\end{equation}

\begin{equation}
=\frac{1}{[G_{kk}+G_{ll}-G_{kl}]}\,.
\end{equation}

Thus

\begin{equation}
R_{kl}=G_{kk}+G_{ll}-G_{kl}.
\end{equation}

\subsection{Gaussian Integral}

Given a matrix $K$ with all positive eigenvalues except for one zero
eigenvalue (with eigenvector $c$) we can define a Gaussian integral

\begin{equation}
Z(J)=\frac{1}{Z}\int_{V}e^{-\frac{1}{2}\sum_{ij}K_{ij}\phi_{i}\phi_{j}+\sum_{i}J_{i}\phi_{i}}d\phi\equiv\langle e^{J\cdot\phi}\rangle
\end{equation}

where $\sum_{i}J_{i}=0$ . The normalization factor $Z$ is chosen
such that $Z(0)=1$.

Also, the range of integration is $V\equiv\mathbb{R}^{n}/\mathbb{R}$
; the quotient of $\mathbb{R}^{n}$ by the translation $\phi_{i}\mapsto\phi_{i}+ac_{i}$.
From each such orbit we can pick a representative that satisfies

\begin{equation}
\sum_{i}\phi_{i}=0.
\end{equation}

This is an elementary example of ``gauge fixing''.

On this $n-1$ dimensional subspace $K$ is invertible with the inverse
$G$ defined above, So

\begin{equation}
Z(J)=e^{\frac{1}{2}\sum_{ij}G_{ij}J_{i}J_{J}}.
\end{equation}

In particular

\begin{equation}
\langle\phi_{k}\phi_{l}\rangle=G_{kl}
\end{equation}

and

\begin{equation}
\langle\left(\phi_{k}-\phi_{k}\right)^{2}\rangle=G_{kk}+G_{ll}-2G_{kl}.
\end{equation}

Thus, the effective resistance is equal to the variance of the voltage
fluctuations:

\begin{equation}
R_{kl}=\langle\left(\phi_{k}-\phi_{k}\right)^{2}\rangle.
\end{equation}

This point of view on the resistance is especially convenient if we
average over $K$ (e.g., percolation). We hope to return to this issue
in another publication.

This procedure for deriving a formula for variance breaks down in
the continuum limit. We need to work not with the potential itself,
but an average of it over a small region.

\section{Abstract Wiener Spaces}

There is another point of view on the regularity of random processes,
based on function spaces. Given an orthonormal basis $e_{n}$ in an
infinite dimensional Hilbert space $H$ we can try to define a random
variable

\begin{equation}
\phi=\sum_{n}g_{n}e_{n}
\end{equation}

where $g_{n}$ are independent Gaussian random variables of zero mean
and variance one. But the probability of this series converging in
the norm of $H$ is zero. For convergence, we need a weaker norm.
More precisely, we seek a Banach space $B$ and an embedding $i:H\to B$
such that the sum converges to a random variable valued in $B$. Such
a triple $(i,H,B)$ is the abstract Wiener Space of Gross \cite{GrossWienerSpace}.
There is no ``best possible'' $B$; the choice is usually motivated
by physics or geometry.

Recall that the Sobolev space $H^{s}$ is the Hilbert space equipped
with inner product $(f,\Delta^{s}g)$. For Brownian motion, the Hilbert
space $H$ defined above is the Sobolev space $H^{1}$ of functions
whose derivatives are square integrable. One choice for $B$ is the
space of continuous functions. A more refined choice would be the
space of functions with norm

\begin{equation}
\sup_{x,y}\frac{|f(x)-f(y)|}{\omega(x,y)},
\end{equation}
where $\omega$ is the L$\mathrm{\acute{e}}$vy modulus described
above. What is the abstract Wiener Space for a massless scalar quantum
field? Note that the ground state wave function of such a field is
(in the notation preferred by physicists) 
\begin{equation}
\psi(\phi)\propto e^{-\frac{1}{2}\int|k||\tilde{\phi}(k)|^{2}\frac{dk}{2\pi}}
\end{equation}
The quadratic form in the exponent can be written as 
\begin{equation}
(\phi,\sqrt{\Delta}\phi)
\end{equation}
where $\Delta$ is the Laplacian and $(f,f)=\int|f(x)|^{2}dx$. Thus,
in more mathematical language, the log-correlated scalar field is
the Gaussian process modeled on the Sobolev space $H^{\frac{1}{2}}(\mathbb{R})$
.

Gross \cite{emails} has shown that any choice of $B$ must fit within
a small band of Hilbert spaces: $L^{2}\subset B\subset H^{-\epsilon}$
for $\epsilon>0$. We can make a proposal for the Abstract Wiener
Space for the massless scalar field on the real line, based on the
modulus of continuity: the completion of the space of continuous functions
(modulo constants) by the norm 
\begin{equation}
||f||\_\omega=\sup_{u,v,s}\frac{|\bar{f}_{s}(u)-\bar{f}_{s}(v)|}{\omega(|u-v|)},\quad\omega(u,v)\approx|u-v|\log\frac{1}{|u-v|}
\end{equation}

\newpage

\providecommand{\noopsort}[1]{}\providecommand{\singleletter}[1]{#1}

\makeatletter \providecommand{\@ifxundefined}[1]{%
 \@ifx{#1\undefined}
}\providecommand{\@ifnum}[1]{%
 \ifnum #1\expandafter \@firstoftwo
 \else \expandafter \@secondoftwo
 \fi
}\providecommand{\@ifx}[1]{%
 \ifx #1\expandafter \@firstoftwo
 \else \expandafter \@secondoftwo
 \fi
}\providecommand{\natexlab}[1]{#1}\providecommand{\enquote}[1]{``#1''}\providecommand{\bibnamefont}[1]{#1}\providecommand{\bibfnamefont}[1]{#1}\providecommand{\citenamefont}[1]{#1}\providecommand{\href@noop}[0]{\@secondoftwo}\providecommand{\href}[0]{\begingroup \@sanitize@url \@href}\providecommand{\@href}[1]{\@@startlink{#1}\@@href}\providecommand{\@@href}[1]{\endgroup#1\@@endlink}\providecommand{\@sanitize@url}[0]{\catcode `\\12\catcode `\$12\catcode
  `\&12\catcode `\#12\catcode `\^12\catcode `\_12\catcode `\%12\relax}\providecommand{\@@startlink}[1]{}\providecommand{\@@endlink}[0]{}\providecommand{\url}[0]{\begingroup\@sanitize@url \@url }\providecommand{\@url}[1]{\endgroup\@href {#1}{\urlprefix }}\providecommand{\urlprefix}[0]{URL }\providecommand{\Eprint}[0]{\href }\providecommand{\doibase}[0]{http://dx.doi.org/}\providecommand{\selectlanguage}[0]{\@gobble}\providecommand{\bibinfo}[0]{\@secondoftwo}\providecommand{\bibfield}[0]{\@secondoftwo}\providecommand{\translation}[1]{[#1]}\providecommand{\BibitemOpen}[0]{}\providecommand{\bibitemStop}[0]{}\providecommand{\bibitemNoStop}[0]{.\EOS\space}\providecommand{\EOS}[0]{\spacefactor3000\relax}\providecommand{\BibitemShut}[1]{\csname bibitem#1\endcsname}\let\auto@bib@innerbib\@empty
%</preamble>

\end{document}